\documentclass{ws-p8-50x6-00}

\begin{document}

\title{Spin structure of the $\Delta(1232)$ \\
       and inelastic Compton scattering}

\author{A.I. L'vov}

\address{
   Lebedev Physical Institute, Leninsky Prospect 53, Moscow 117924, Russia \\
   E-mail: lvov@x4u.lebedev.ru}

\maketitle

\abstracts
{Radiative transitions $\gamma\Delta(1232)\to N^*$ are discussed in the
nonrelativistic quark model with spin-orbit corrections for the 70-plet
$L^P=1^-$ nucleon resonances $N^*$. The reaction $\gamma N\to\gamma\Delta$
is considered as a tool to measure some of these transitions.
A particular sensitivity to photoexcitations of $S_{11}(1535)$,
$D_{13}(1700)$, and $D_{15}(1675)$ is predicted.}

\section{Introduction: the GDH sum rule for the $\Delta(1232)$}

My motivation to analyze photocouplings of the $\Delta(1232)$ was
initially related to a problem of saturation of the
Gerasimov--Drell--Hearn sum rule \cite{gerasimov65} for the $\Delta$
\cite{lvov00}. Generally, the GDH integral
\begin{eqnarray}
\label{GDH}
  I &=& \int_{\rm thr}^\infty \frac{d\omega}{\omega}
    \left[ \sigma_{1+s}(\omega) - \sigma_{1-s}(\omega) \right]
    = \pi s \left( \frac{\mu}{s} - \frac{eZ}{M} \right)^2
\\
\label{GDH-res}
    &=& \sum_{N^* \ne B} \frac{4\pi M^* M}{(M^{*\,2} - M^2)^2}
     \left\{ |A(1,s)|^2 - |A(1,-s)|^2 \right\},
\end{eqnarray}
must give the magnetic moment $\mu$ of the baryon $B$ of the mass $M$,
spin $s$ and the electric charge $eZ$ through the total photoabsorption
cross sections $\sigma_{1\pm s}$ for the target spin parallel or
anti-parallel to the photon helicity $\lambda=+1$.  The second line
re-expresses the integral (\ref{GDH}) through the electromagnetic
transition amplitudes $A(\lambda,\sigma)$ of $\gamma B\to N^*$
in the zero-width approximation for the resonances $N^*$. Usually
photocouplings $A_\Lambda$ are defined instead of $A(\lambda,\sigma)$,
\begin{equation}
   A(\lambda,\sigma) \equiv \langle N^*(\vec p+\vec k), \lambda+\sigma \mid
    {-\vec\epsilon_\lambda}\cdot \vec J_{\rm e.m.}(\vec k)
      \mid B(\vec p), \sigma \rangle
     = \sqrt{\frac{M^{*\,2} - M^2}{M^*}} \,\xi A_\Lambda \,,
\label{Arad}
\end{equation}
where $\Lambda=\lambda+\sigma$ is the total helicity, $\lambda=+1$, and
$\xi=\pm 1$ is a phase factor determined by the pion decay amplitude of
the resonance $N^*$~\cite{capstick00}.  Collinear kinematics $\vec p
\parallel \vec k$ is assumed in (\ref{Arad}).

In the framework of the nonrelativistic quark model (NQM), there is a
fundamental difference between the GDH sum rule for the nucleon and
for the $\Delta$. The radiative $M1$ transitions $N\to\Delta$ with the
helicities $1+\frac12 \to \frac32$ and $1+(-\frac12)\to \frac12$ nearly
saturate the sum rule for the nucleon giving $I=216~\mu$b in close
agreement with the experimental values $I_p=205~\mu$b and
$I_n=233~\mu$b inferred from the nucleon anomalous magnetic moments.
Meanwhile the same $N\leftarrow\Delta$ transition gives a big {\em
negative} contribution $I=-323~\mu$b to the GDH sum rule for the
$\Delta$ [note the helicities $1+(\frac{-3}2) \to -\frac12$] in drastic
disagreement with the manifestly positive r.h.s.\ of Eq.~(\ref{GDH}).
Assuming that the GDH sum rule is valid for all hadrons including the
$\Delta$, we conclude that the $\Delta$ must have strong photocouplings
to nucleon resonances of spin $J \ge \frac52$.  These resonances must
compensate $\Delta\to N,N^*$ contributions from spin $J \le \frac32$
states, all of which are negative.

\section{Photocouplings in NQM: the role of spin-orbit corrections}

It was found long ago \cite{close70} that nonrelativistic constituent
models do not obey the GDH constraint (\ref{GDH}).  Relativistic
corrections are needed in order to restore the validity of the GDH sum
rule. They are especially important for high-lying states.  For a
target having all constituents (quarks) on the $s$-shell and thus
having the angular and magnetic moment completely composed of
constituent's spins, it is sufficient to include only spin-orbit
corrections,
\begin{equation}
   H_{\rm e.m.}^{\rm SO+NA} = \sum_i \frac{e_i\vec\sigma_i}{8m_i^2} \,
        \cdot(\vec p_i\times\vec E_i - \vec E_i\times\vec p_i)
   + \sum_{i\ne j} \frac{e_i}{4M_{\rm tot}} \!
        \left(\frac{\vec\sigma_i}{m_i} - \frac{\vec\sigma_j}{m_j} \right)
       \cdot \vec p_j\times\vec E_i \, .
\label{H-so}
\end{equation}
Apart from ordinary additive spin-orbit terms (SO) this equation
contains important non-additive (NA) two-body pieces which appear owing
to the Wigner rotation of quark spins into the center-of-mass frame of
the baryon \cite{close70,close90}.

One can illustrate the role of the SO$+$NA terms taking the radiative
transition $\gamma N\to\Delta$ as example. Equation (\ref{Arad})
defines the photocoupling $A_\Lambda$ of the nucleon as a
Lorentz-invariant quantity.  However, when nonrelativistic wave
functions and the nonrelativistic electromagnetic current are used in
(\ref{Arad}), the result does depend on the frame used. For example,
the photocoupling $A_{3/2}$ found in the c.m.\ frame ($\vec p=-\vec
k$), in the Breit frame ($\vec p=-\frac12\vec k$) and in the lab frame
($\vec p=0$) is equal to $-175$, $-194$ and $-219\times 10^{-3}~\rm
GeV^{-1/2}$, respectively [oscillator wave functions were used in this
calculation with the oscillator parameter $a=0.41$ GeV and $m_q=336$
MeV].  When the spin-orbit terms (\ref{H-so}) are included, the result
turns out be nearly frame-independent:  $-198$, $-194$ and $-182\times
10^{-3}~\rm GeV^{-1/2}$, respectively.  In this example SO and NA
corrections are equally important and they are seen to nearly restore
the Lorentz invariance of the transition amplitude.

Using the Karl--Isgur quark model (for a recent review see
\cite{capstick00}) with the spin-orbit correction (\ref{H-so}), we
(re)calculated photocouplings of the nucleon to negative-parity baryons
$|70,~L^P=1^-\rangle$. With three exceptions, we found a full agreement
with the previous work~\cite{close90} [those exceptions are the SO
contribution to the amplitude $A_{3/2}$ for $\gamma N\to D_{33}(1700)$
and NA contributions to the amplitudes $A_{1/2}$, $A_{3/2}$ for $\gamma
N\to S_{13}(1620)$].

Then we calculated photocouplings of the $\Delta$. Our results for
$\Delta^+$ are given in the Table~\ref{tab} as example, separately for
the cases when a pure nonrelativistic approximation is used (columns
NR) and when the spin-orbit corrections (\ref{H-so}) are included too
(columns SO). Also (some of) predictions of Carlson--Carone
\cite{carlson98} are shown (columns CC1) which have been obtained
through a fit to experimental data on {\em nucleon} photocouplings
using one-body quark operators dominating in the large $N_c$ limit of
QCD.  For {\em large} amplitudes, there is a qualitative agreement
between our results (when SO is included) and the results of
Carlson--Carone with the except for a few signs which are probably
related with different phase conventions used for the baryon wave
functions; our convention follows Ref.~\cite{capstick00}).

Our calculation suggests a strong photocoupling of the $\Delta(1232)$
to the lowest $J=\frac52$ state which is $D_{15}(1675)$.  The
$\Delta\to D_{15}$ transition contributes much to the GDH sum rule for
the $\Delta$ and it helps to reduce a gap between the negative
Eq.~(\ref{GDH-res}) evaluated through the lowest 56 and 70-plet baryons
$N^*$ and the manifestly positive r.h.s.\ of Eq.~(\ref{GDH}).  See
\cite{lvov00} for further detail.

\begin{table}[hbt]
\caption{$\Delta^+(1232)$ photocouplings to $|70,~L^P=1^-\rangle$
baryons in units of $10^{-3}~\rm GeV^{-1/2}$. The phase factor of
$\xi$, Eq.~(\protect\ref{Arad}), is included into $A_\Lambda$.}
\label{tab}
\small
$
\begin{array}{|c|rrr|rrr|rrr|rrr|}
\hline
   N^*    && A_{-1/2} &&& A_{1/2} &&& A_{3/2} &&& A_{5/2} & \\
\cline{2-13}
& \rm NR & \rm SO & \rm CC1
& \rm NR & \rm SO & \rm CC1
& \rm NR & \rm SO & \rm CC1
& \rm NR & \rm SO & \rm CC1
\\
\hline
S_{11}(1535) & -88 & -75 & 108 & -94 & -86 &   4 &     &     &     &&& \\
D_{13}(1520) &  61 &  20 &  62 &  48 &   1 & -16 &  22 & -19 & -90 &&& \\
S_{11}(1650) &-196 &-115 & -63 &-107 & -60 &-128 &     &     &     &&& \\
D_{13}(1700) & -87 & -64 &  43 &-164 &-138 & 123 &-198 &-175 & 169 &&& \\
D_{15}(1675) &   9 & -12 &  24 & -26 & -62 & -19 & -97 &-147 &-113 &-201 &-267 &-258 \\
S_{31}(1620) &  46 &  46 & -42 & -27 & -27 &  73 &     &     &     &&& \\
D_{33}(1700) &  81 &  81 & -54 &  47 &  47 &   0 &   0 &   0 &  55 &&& \\
\hline
\end{array}
$
\end{table}

\section{Differential cross section of $\gamma N\to \gamma\Delta$}

\begin{figure}[hbt]
\centerline{\epsfxsize=0.7\textwidth\epsfbox{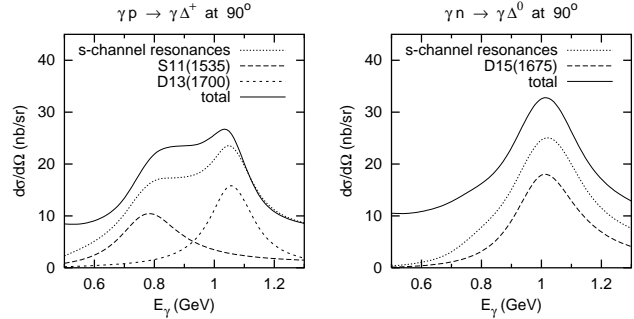}}
\caption{Differential cross sections (c.m.) of inelastic
Compton scattering in the NQM including spin-orbit corrections.
Contributions of $s$-channel resonances (without $u$-channel and
seagull contributions) are shown separately.}
\label{fig}
\end{figure}

Using the above photoamplitudes as functions of the photon energy, we
estimated differential cross section of inelastic Compton scattering
$\gamma N\to \gamma\Delta$ in the resonance region. Both $s$ and $u$
channel resonances, with experimental masses and widths, were included
as well as seagull contributions, cf.\ Ref.~\cite{capstick92}.
Dominating contributions to the reaction off the proton come
from the $S_{11}(1535)$ and $D_{13}(1700)$ resonances, see
Fig.~\ref{fig}. For the neutron target, the $D_{11}(1675)$ resonance
dominates.  Note that $D_{11}(1675)$ is decoupled from the proton [i.e.,
the Moorhouse selection rule remains valid for $D_{15}$
even after the NA corrections are included], so that the reaction off
the neutron, via $\gamma d\to \gamma p_s \Delta^0$, turns out the only
efficient way to study the $\gamma\Delta D_{15}$ transition.

The differential cross section of inelastic Compton scattering is seen
to be compatible with that of elastic $\gamma N$ scattering.  Therefore
we may conclude that the original idea of Carlson--Carone
\cite{carlson98} to study photocouplings of $\Delta$ in the reaction
$\gamma N\to \gamma\Delta$ can probably be realized in practice.

\end{document}